\documentclass[aps, prd, amsmath, amssymb, twocolumn, showpacs]{revtex4}
\usepackage{graphicx}
\usepackage{color}
\usepackage{natbib}
\begin{document}


\title{The Self-Gravity of Pressure in Neutron Stars}
\author{Josiah Schwab}
\author{Scott A.\ Hughes}
\author{Saul Rappaport}
\affiliation{Department of Physics and Kavli Institute for
Astrophysics and Space Research, MIT, Cambridge, MA 02139}


\begin{abstract}

Following an earlier analysis which examined the effect of the
self-gravity of pressure on big-bang nucleosynthesis (BBN), we explore
the effect of pressure's self-gravity on the structure of neutron
stars.  We construct an {\em ad hoc} modification of the
Tolman-Oppenheimer-Volkoff equation wherein pressure's self-gravity is
parameterized by a constant, $\chi$, with $0 \le \chi \le 1$.  The
full general relativistic contribution to the gravity of pressure is
recovered with $\chi = 1$, and is eliminated when $\chi = 0$.  This
formulation is not proposed as an alternative theory of gravity, but
is merely used to quantify the extent to which the self-gravity of
pressure contributes to the structure of dense objects.  As can be
surmised qualitatively, neutron star masses can be quite sensitive to
$\chi$, with higher values of neutron-star mass (by $\sim$20--25\%)
allowed for smaller values of $\chi$.  However, for a given equation
of state, neither the range of neutron star radii nor the radii at fixed central 
density depend
sensitively on $\chi$.  Over the neutron star mass range measured so
far, the presence or absence of pressure's self-gravity yields a
nearly immeasurable change in radius --- much smaller than the
variations in radius due to the uncertainty in the equation of state.
In contrast to the result for BBN, we thus find that neutron stars are
not likely to be useful testbeds for examining the self-gravity of
pressure.

\end{abstract}



\maketitle


\section{Introduction}

One of the more fascinating predictions of general relativity is that
pressure is self gravitating.  This is a strictly non-Newtonian
result, whereby the pressure of a field contributes to gravity,
increasing the effective density by an amount $3 P/c^2$ for a
homogeneous perfect fluid.

There are four prominent physical situations in which the self-gravity
of pressure could potentially lead to a measurable effect (three of
which involve the expansion history of the Universe): (i) the
acceleration of the Universe during inflation; (ii) the expansion of
the Universe during the radiation-dominated epoch; (iii) the current
acceleration of the Universe due to dark energy; and (iv) the
mass-radius relation for neutron stars.

In an earlier paper, Rappaport et al.\ \cite{Rappaport08} addressed
(ii) by setting a constraint on the self-gravity of pressure during
the radiation-dominated epoch of big bang nucleosynthesis (BBN). They
introduced an {\it ad hoc} multiplicative parameter $\chi$ to the $3
P/c^2$ contribution in the Einstein field equations.  For $\chi = 1$,
the full general relativistic contribution to the self-gravity of
pressure is retained; for $\chi = 0$, there is no such contribution.
This yielded a modified set of Friedmann-like equations.  The
contribution to $(\dot a/a)^2$ for a species of matter with equation
of state (EOS) $w = P/\rho c^2$ appears as
\begin{equation}
\left(\frac{\dot a}{a}\right)^2 =
\left(\frac{1+3w\chi}{1+3w}\right)\frac{\Omega}{a^{1+3w}}\;.
\label{eq:scaling}
\end{equation}
The gravitational effect of radiation ($w = 1/3$) is therefore scaled
by $(1+\chi)/2$.  The absence of presure's self-gravity means that the
Hubble constant at the time of BBN would be smaller by a factor of
$\sqrt{2}$, potentially changing the abundance of light elements.
Rappaport et al.\ used a standard BBN code and performed light element
abundance calculations for a range of $\chi$ to quantify this effect.
When combined with current light element observations, the data were
shown to be consistent with $\chi = 1$, and strongly exclude $\chi =
0$

Our goal here is to see whether similar limits can be set by
observations of neutron stars.  At neutron star densities, $P \sim
(0.1 - 0.6)\rho c^2$, suggesting that pressure is high enough that a
measureable self-gravity effect might exist.  We introduce a
parameterization of self-gravity very similar to that used to modify
the Friedmann equations.  The result is a parameterized equation of
stellar structure, with $\chi = 1$ reproducing the
Tolman-Oppenheimer-Volkoff equation, and $\chi = 0$ ``turning off''
the pressure self-gravity of that equation.

Although at fixed central density $\chi$ has a significant impact on a
neutron star's mass, it has very little impact on its radius.  If
measurements were to determine both mass and radius, it would still be
extremely difficult to tell the difference between models with $\chi =
0$ and $\chi = 1$ due to the uncertainty in the neutron star EOS.
Neutron stars thus appear to {\it not} be very
useful for testing the self-gravity of pressure.  This is not to say
that one cannot make interesting statements about gravity with neutron
stars {\cite{psaltis07}}; but, in contrast to the situation with BBN,
the pressure self-gravity aspect cannot be usefully tested.  The
difference between the two cases is simple: the EOS of the universe during
BBN is well understood, but the EOS of neutron stars is not.  Testing
pressure's self gravity is thus degenerate with testing the EOS for
neutron stars, but is not degenerate during BBN.

\section{Structure of Neutron Stars}

In general relativity, the structure of spherically symmetric fluid
equilibria such as neutron stars is governed by the
Tolman-Oppenheimer-Volkoff (TOV) equation \cite{Tolman34,
Oppenheimer39}.  To motivate our heuristic, it is useful to review the
derivation of the TOV equation.  Further details can be found in, for
example, Ref.\ {\cite{MTW}}, Sec.\ 23.5.

We begin by assuming static, spherical symmetry in a perfect fluid.
The perfect fluid stress-energy tensor is
\begin{equation}
T_{\mu\nu} = (\rho + P)u_\mu u_\nu + P g_{\mu\nu}\;,
\label{eq:tmunu}
\end{equation}
where $\rho$ is the fluid's local energy density, $P$ its pressure
(with $c = 1$), and $u_\mu$ is a component of its 4-velocity.  In
equilibrium, the fluid is static so its only non-zero component is
$u_t$.  The metric follows from the line element
\begin{equation}
ds^2 = -e^{2\Phi} \ dt^2 + e^{2\Lambda} \ dr^2 + r^2 \ d\theta^2 +
r^2\sin^2\theta \ d\phi^2\;.
\label{eq:metric}
\end{equation}
The condition $g_{\mu\nu} u^\mu u^\nu = -1$ sets $u_t = -e^{\Phi}$.
From the metric, we compute the Einstein tensor $G_{\mu\nu}$ and
enforce the Einstein field equation (EFE) $G_{\mu\nu} = 8\pi G
T_{\mu\nu}$:
\begin{eqnarray}
G_{tt}  = & 
\frac{1}{r^2}e^{2(\Phi - \Lambda)}\left(2r\frac{\partial \Lambda}{\partial r} - 1 + e^{2\Lambda}\right)
& =  8\pi G e^{2\Phi} \rho
\label{eq:efe_tt}\\
G_{rr}  = & 
\frac{1}{r^2}\left(2r\frac{\partial \Phi}{\partial r} + 1 - e^{2\Lambda}\right)
& =  8\pi G e^{2\Lambda} P\;.
\label{eq:efe_rr}
\end{eqnarray}
Because of spherical symmetry, the $G_{\theta \theta}$ and
$G_{\phi\phi}$ components do not contain any additional information.

Local conservation of energy, expressed as $\nabla_{\mu} T^{\mu\nu} =
0$ (where $\nabla_\mu$ denotes the covariant derivative), gives
\begin{equation}
\frac{dP}{dr} = -(\rho + P) \frac{d\Phi}{dr}\;.
\label{eq:econs2}
\end{equation}
Note that $(\rho + P)$ plays the role of an inertial mass density
here.  It sets the proportionality for the force a fluid element must
feel in order not to experience free fall.

With a little rearranging and a judicious definition, Eq.\
(\ref{eq:efe_tt}) describes how the star's gravitational mass
accumulates as a function of radius:
\begin{equation}
m(r) \equiv \frac{r}{2}\left(1 - e^{-2\Lambda}\right)
= 4\pi \int_0^r \rho(r')(r')^2 dr'\;.
\label{eq:massdef}
\end{equation}
This result may appear intuitively obvious.  Note, however, that $4\pi
r^2dr$ is not the {\it proper} spherical spatial volume in the metric
(\ref{eq:metric}); it lacks a factor of $\sqrt{g_{rr}}$.  The
difference between $m(r)$ and a mass $m'(r)$ that includes this factor
can be interpreted as the gravitational binding energy of the star.

Combining Eqs.\,(\ref{eq:massdef}) and (\ref{eq:efe_rr}) into
Eq.\,(\ref{eq:econs2}) at last yields
\begin{equation}
\frac{dP}{dr} = -(\rho + P) \frac{G[m(r) + 3P (4\pi r^3/3)]}{r^2 [1 -
2Gm(r)/r]}\;.
\label{eq:TOV}
\end{equation}
This is the standard form of the TOV equation.  Note the varying roles
$P$ plays here.  The $\rho + P$ term acts as an inertial mass density.
By contrast, $m(r) + 3P (4\pi r^3/3)$ acts as a gravitational mass. In
any metric theory of gravity, this mass distinction arises because the
$dP/dr$ term represents a force accelerating a fluid element with
inertial mass density $(\rho + P)$ away from its geodesic.

Thus, only the second $P$ term in Eq.\ (\ref{eq:TOV}) represents the
gravitating component of pressure.  We now parameterize the 
contribution of pressure to $dP/dr$
by modifying the TOV equation as follows:
\begin{equation}
\frac{dP}{dr} = -(\rho + P) \frac{G[m(r) + 3P \chi(4\pi r^3/3)]}{r^2
[1 - 2Gm(r)/r]}\;.
\label{eq:TOVmod}
\end{equation}
In contrast to the modified Friedmann equations discussed in the
Introduction, $\chi = 0$ does not correspond to a Newtonian limit.
The $[1 - 2Gm(r)/r]$ term in the denominator is a statement of
geometry, and the $(\rho + P)$ term arises from conservation of energy
that must be valid even in the special relativity limit.

It is worth emphasizing that Eq.\ (\ref{eq:TOVmod}) is {\it not}
derived by proposing a plausible alternative theory of gravity.  Our
goal is much less ambitious than reformulating gravity; we merely want
to quantify the extent to which the self-gravity of pressure
contributes to the structure of dense objects.  The point of this
exercise was to see which of the pressure terms in Eq.\ (\ref{eq:TOV})
acts as a source of gravity, and then to appropriately modify that
equation in a way that tests that term's importance.

\section{Model building, results, and discussion}

The first step in building neutron star models is to select an EOS,
$\rho = \rho(P)$.  For high densities, we use the compilation of
Lattimer \& Prakash \cite{Lattimer01}, plus two of the ``mixed'' EOSs
from Alford et al.\ \cite{Alford05}. Our set includes a wide range of
models, representing several different theoretical approaches, which incorporate both normal
nuclear matter compositions and more exotic matter (quarks, hyperons,
etc.).  Table \ref{tbl:eos} gives an EOS summary, listing general
properties and giving relevant references.  The trailing numerical labels are
as given in \cite{Lattimer01}.  For the ``ALF'' EOSs, the numerical label
indicates the transition density in multiples of the nuclear
saturation density. In all cases, we use the Negele-Vautherin (NV) EOS
\cite{Negele73} for the lower baryon density range $0.08 > n_b > 6
\times 10^{-4}$ fm$^{-3}$, and we use the Baym-Pethick-Sutherland
(BPS) EOS \cite{BPS71} for $n_b < 6 \times 10^{-4}$ fm$^{-3}$.  This is similar to
the low density choices made in \cite{Lattimer01}.

\begin{table*}[ht]
\begin{ruledtabular}
\begin{tabular}{lllc}
Symbol & Reference & Approach & Composition \\\hline
AP(3-4)	& Akmal \& Pandharipande \cite{Akmal97}	& Variational &	np \\
MPA & M\"{u}ther, Prakash, \& Ainsworth \cite{Muther87} & Dirac-Brueckner HF & np \\
MS(1-3)	& M\"{u}ller \& Serot \cite{Muller96} & Field Theoretical & np \\
WFF(1-3) & Wiringa, Fiks, \& Fabrocine \cite{Wiringa88}	& Variational &	np \\
\hline
ALF(2-3) & Alford et al. \cite{Alford05} & Quark Matter & npQ \\
GM(1-3)	& Glendenning \& Moszkowski \cite{Glendenning91} & Field Theoretical & npH \\
GS(1-2)	& Glendenning \& Schaffner-Bielich \cite{Glendenning99} & Field Theoretical & npK \\
PCL & Prakash, Cooke, \& Lattimer \cite{Prakash95} & Field Theoretical & npHQ \\
\end{tabular}
\end{ruledtabular}

\caption{Summary of the EOSs shown in Figure \ref{fig:NS-MR} and
\ref{fig:two-shade}.  ``Approach'' characterizes the theoretical
basis.  ``Composition'' is characterized by (n -- neutrons, p --
protons, H -- hyperons, K -- kaons, Q -- quarks). The horizontal line
separates ``normal'' and ``exotic'' equations of state.
\label{tbl:eos}}
\end{table*}

We construct the neutron star models by numerically solving the
coupled differential equations
\begin{eqnarray}
\frac{d\rho}{dr} &= &\frac{d\rho}{dP}\frac{dP}{dr}
\label{eq:drhodr}\\
\frac{dM}{dr} &=& 4\pi r^2 \rho\;.
\label{eq:dMdr}
\end{eqnarray}
The tabulated EOSs were logarithmically interpolated and the
derivatives evaluated by a simple 3-point Lagrange interpolation.
Integration proceeds until the density falls below that of iron, 7.9 g
cm$^{-3}$. We constructed models with both $\chi = 0$ and $\chi = 1$
for each EOS.

Figure 1 illustrates the effect that varying $\chi$ has for a
particular EOS, AP4.  We show the mass-radius relation for $\chi = 1$
(black curve) and $\chi = 0$ (red curve).  Notice that the radial
range spanned by these two models is largely the same; the mass,
however, can change substantially.  The arrows in this figure connect
models that have equal central densities.  This shows that pressure's
self-gravity can have a $\sim$10 -- 30\% effect on the mass, but
barely change the star's radius---at a fixed central density.  
One might have guessed on intuitive
grounds that an effect of order tens of percent must occur, since for
these models $P/\rho \sim$ 0.1 -- 0.6.  The fact that this change is
mostly in mass, leaving the radius relatively unaffected, was not 
obvious in advance. Finally, in this regard, we note that for a given EOS and
at a fixed neutron-star mass near the upper mass limit, the radius can
vary by up to $\sim$15\% when $\chi$ is reduced to near zero.  However,
the differences in radii at a fixed mass, among the various plausible
EOSs, are much larger than this, thereby making it difficult, at best, to
constrain $\chi$

\begin{figure}[ht]
\includegraphics[width = 0.47 \textwidth]{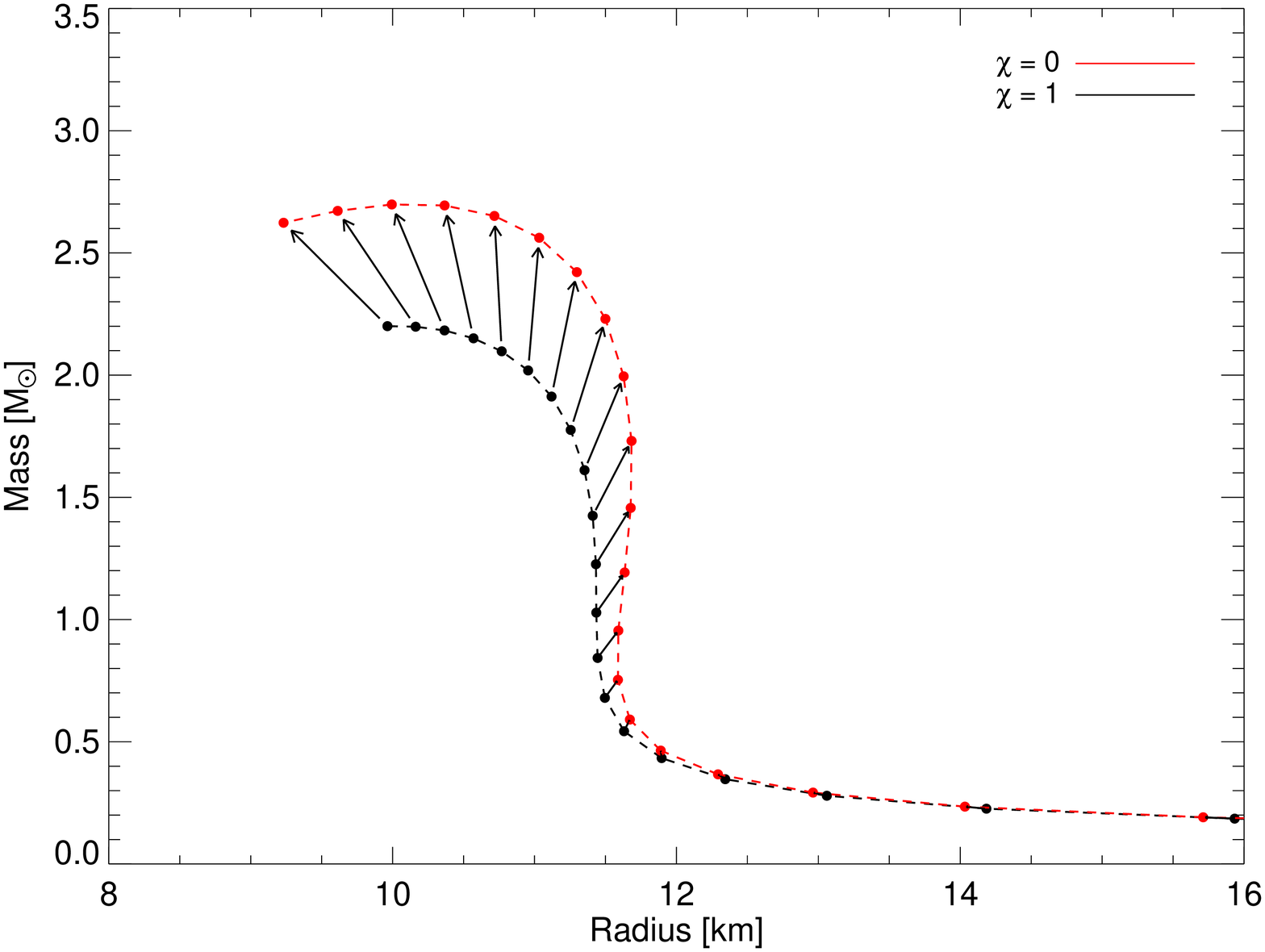}
\caption{For a representative EOS (AP4), the mass-radius relations for
GR ($\chi = 1$, black) and without the self-gravity of pressure ($\chi
= 0$, red) are compared. For the $\chi = 1$ and $\chi = 0$ curves we
connect models with equal central density.  We note that the last two
arrows on the left side of the plot point from stable $\chi = 1$
models to {\it unstable} $\chi = 0$ models.  Thus, there are no stable
models where the radius changes by more than $\sim5\%$ due to
variation in $\chi$ for this particular EOS. \label{fig:two-chi}}
\end{figure}

\begin{figure}[ht]
\includegraphics[width = 0.47 \textwidth]{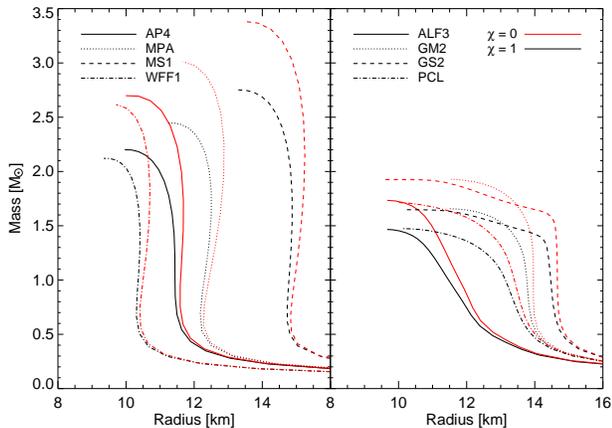}
\caption{Mass-radius relations for a variety of EOSs. The general
relativistic result ($\chi = 1$) is shown in black, while the results
in the absence of the self-gravity of pressure ($\chi = 0$) are shown
in red. The left panel shows EOSs of nuclear matter. The right panel
contains equations with more exotic kinds of matter.
\label{fig:NS-MR}}
\end{figure}

Figure \ref{fig:NS-MR} shows the effect of removing the self-gravity
of pressure for several illustrative EOSs.  The general trend seen in
Fig.\ {\ref{fig:two-chi}} --- an increase in mass for a given radius
--- holds for both ``normal'' (left panel) and ``exotic'' (right
panel) compositions.  In the most compact stars, $\chi = 0$
corresponds to a change of $\sim 0.5 M_\odot$ over the nominal $\chi =
1$ model, for a constant radius.

In the case of BBN, the cosmic abundance of light elements puts
interesting constraints on $\chi$.  Might it be possible to place
similar limits on $\chi$ from neutron star observations?  The answer,
unfortunately, appears to be ``no''.  Neutron star masses have in
some cases been measured to accuracies of greater than 1 part in 1000,
most notably PSR 1913+16 \cite{Weisberg05}), where the pulsar and its
companion have measured masses of $1.441 M_\odot$ and $1.387 M_\odot$
respectively.  A measurement of the radius of one of these objects
would place strong constraints on the EOS.  Referring to Fig.\
{\ref{fig:NS-MR}}, in this mass range and for ``normal'' EOSs, it
would take a radius measurement with an accuracy $\sim$300 m in order
to, for a given EOS, differentiate between the $\chi = 0$ and $\chi =
1$ cases.  A number of proposed techniques for measuring neutron star
radii have been recently summarized in \cite{Lattimer07}.  While
inventive and intriguing, none of these techniques has yet led to
robust determinations of neutron-star radii.

\begin{figure}[ht]
\includegraphics[width = 0.47 \textwidth]{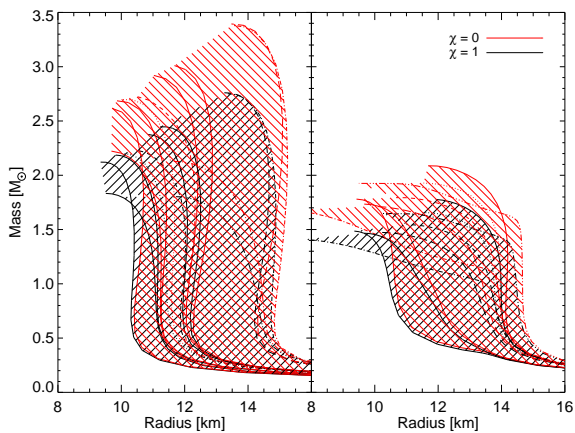}
\caption{Mass-radius relations for a spectrum of EOSs.  The shaded
area fills the region bounded by the most extreme M(R) curves.  The
general relativistic result ($\chi = 1$) is shown in black, while the
results in the absence of the self-gravity of pressure ($\chi = 0$)
are shown in red. \label{fig:two-shade}}
\end{figure}

Figure \ref{fig:two-shade} makes this point even more clearly,
including a more comprehensive sample (all EOSs listed in Table 1) and
showing the ranges of possible masses and radii as shaded regions,
bounded by the most extreme EOSs.  The change in the region between
the $\chi = 0$ and $\chi = 1$ cases clearly shows that the absence of
gravitating pressure acts to stretch possible masses upward, coupled
with a minimal increase in radius.

On this plot, the only plausible evidence for a model with $\chi \ne
1$ would be a precise measurement of both mass and radius that placed
a star in a red-shaded region that is not also a black-shaded region.
For neutron stars with $M \lesssim 2\,M_\odot$ it would be extremely
difficult to distinguish between viable EOSs on the one hand, and the
effect of pressure's self-gravity on the other.  For higher masses,
$2\,M_\odot \lesssim M \lesssim 2.7\,M_\odot$, a measurement of radius
could in principle distinguish the effect of self-gravity from EOS
dependence, at least for those EOSs we have considered here.  Of
course, if any neutron stars are ever found with masses exceeding
$\sim$2 $M_\odot$, the entire set of EOSs would certainly be robustly
reexamined before considering that the equations of general relativity
should be modified to remove the self-gravity of pressure.

Given the difficulty of interpreting such a measurement as a test of
gravity, we conclude that neutron stars (in contrast to BBN) are not a
good laboratory for examining the self-gravity of pressure.


\begin{acknowledgments}
JS acknowledges support from the Paul E. Gray (1954) Endowed Fund for
UROP; SAH is supported by NSF Grant No.\ PHY-0449884 and the MIT Class
of 1956 Career Development Fund.  We thank Krishna Rajagopal for
extremely helpful discussions. We also thank James Lattimer, Madappa
Prakash, and Mark Alford for generously providing equations of state
in a tabulated form.
\end{acknowledgments}



\end{document}